\documentclass[twocolumn,3p]{elsarticle} 

\usepackage{xcolor}
\usepackage{amsmath}
\usepackage{amssymb}
\usepackage{lineno,hyperref}
\usepackage{siunitx}
\usepackage{graphicx}
\graphicspath{{./}{./figures/}} 

\DeclareSIUnit\amp{\ampere}
\DeclareSIUnit\year{yr}
\DeclareSIUnit\mil{mil}

\journal{Nuclear Instruments and Methods}

\begin{document}

\begin{frontmatter}

\title{An In-Situ Movable Calibration Source for Cryogenic Particle Detectors}

\author{N.~Mast}
\author{M.~Fritts} 
\author{D.J.~Sincavage} 
\author{P.~Cushman}
\author{V.~Mandic}

\address{School of Physics \& Astronomy, University of Minnesota, Minneapolis, Minnesota 55455, USA}


\begin{abstract}
A prototype device capable of moving a radioactive calibration source to multiple positions was operated at millikelvin temperatures using a modified commercial stepper motor. It was developed as an in-situ calibration strategy for cryogenic dark matter detectors. Data taken by scanning a calibration source across multiple radial positions of a prototype dark matter detector demonstrated its functionality. Construction, heat load, and operation of the device are discussed, as is the effect of the motor on the detector operation. A sample dataset taken over multiple positions of a SuperCDMS detector is presented as an example of the utility of such a device.
\end{abstract}

\end{frontmatter}



\section{\label{sec:intro}Introduction}
Semiconductor particle detectors can exhibit different responses depending on the location of particle interaction within the detecting medium. This can be the result of surface effects, variations in the material itself, or details of bias electrode arrangements. Position-dependent response is often beneficial in distinguishing surface event backgrounds from bulk interactions \cite{Agnese13,Broniatowski09}. Understanding the position dependence is important for constructing and calculating the associated efficiencies of fiducialization criteria. It is thus desirable to characterize how the detector response depends on location of an event in the detector volume in order to correct for such effects. This can be particularly difficult to measure directly in cryogenic detectors due to the limited access to such devices within a cryostat. 

This work presents an in-situ, movable calibration source, constructed with an inexpensive commercial stepper motor, which can translate a radioactive calibration source across the surface of a detector. With the current design, moving the source presented some difficulties caused by significant heating effects, but a working operational mode was found, demonstrating the promise of the basic design concept.

\section{\label{sec:setup}Setup}
In a procedure similar to that described in \cite{Porter94}, a PF25-48P4 stepper motor was purchased unassembled from Nippon Pulse America Inc., and the coils were rewound with Nb-Ti superconducting wire. Each coil was wound with 900 turns of \SI{4.1}{\mil} (\SI{3}{\mil} core, formvar-clad) Nb-Ti wire from Supercon, Inc. Each coil had a room temperature resistance of \SI{7.5}{\kilo\ohm}. Lubricants on the motor parts were removed by cleaning with methanol in an ultrasonic cleaner bath. The motor was controlled by an EasyDriver \cite{EasyDriver} stepper motor controller and an Arduino Uno \cite{Arduino} microcontroller board. The motor was operated in full-step mode, giving 48 steps per rotation.

\begin{figure}[h]
\includegraphics[width=\columnwidth]{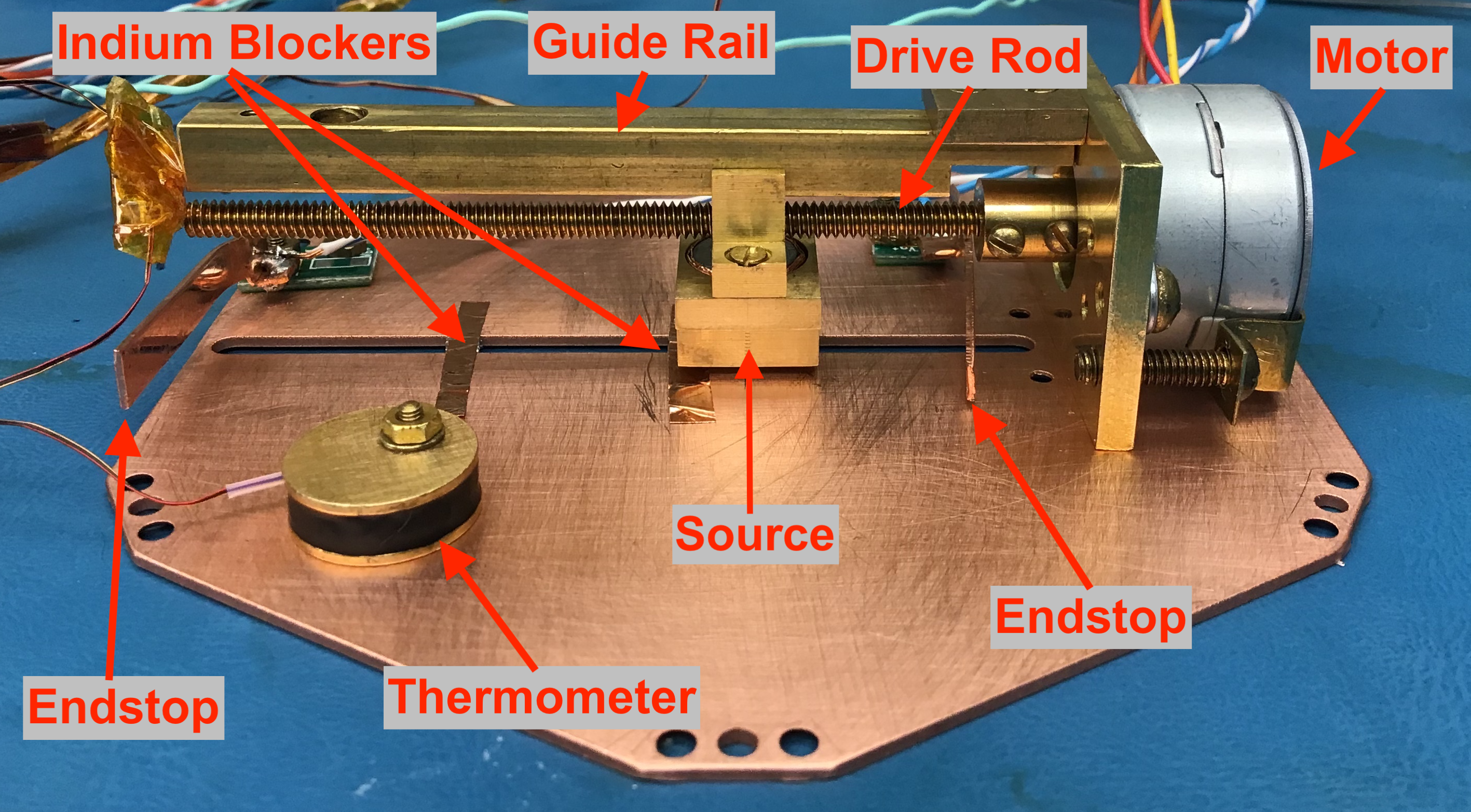}
\caption{\label{fig:motor} Stepper motor and source holder assembly.}
\end{figure}

A brass \#6-32 threaded rod, linked to the motor, drove a source holder equipped with a small radioactive calibration source. As shown in Fig.~\ref{fig:motor} the source holder could be threaded onto the drive rod and fit over a guide rail which caused the motor rotation to linearly actuate the source position. With a count of 32 threads per inch and a motor step angle of \SI{7.5}{\degree}, a single step moved the source a distance of \SI{16.54}{\micro\metre}.

The calibration source used in this experiment was Am-241, with a primary gamma emission energy of 60 keV. A \SI{66}{\mil} thick lead disk with an \SI{18}{\mil} hole collimated the source gammas and restricted the emission rate. A strip of Kapton tape placed over the collimator blocked alpha emission.

\begin{figure}[h]
\includegraphics[width=\columnwidth]{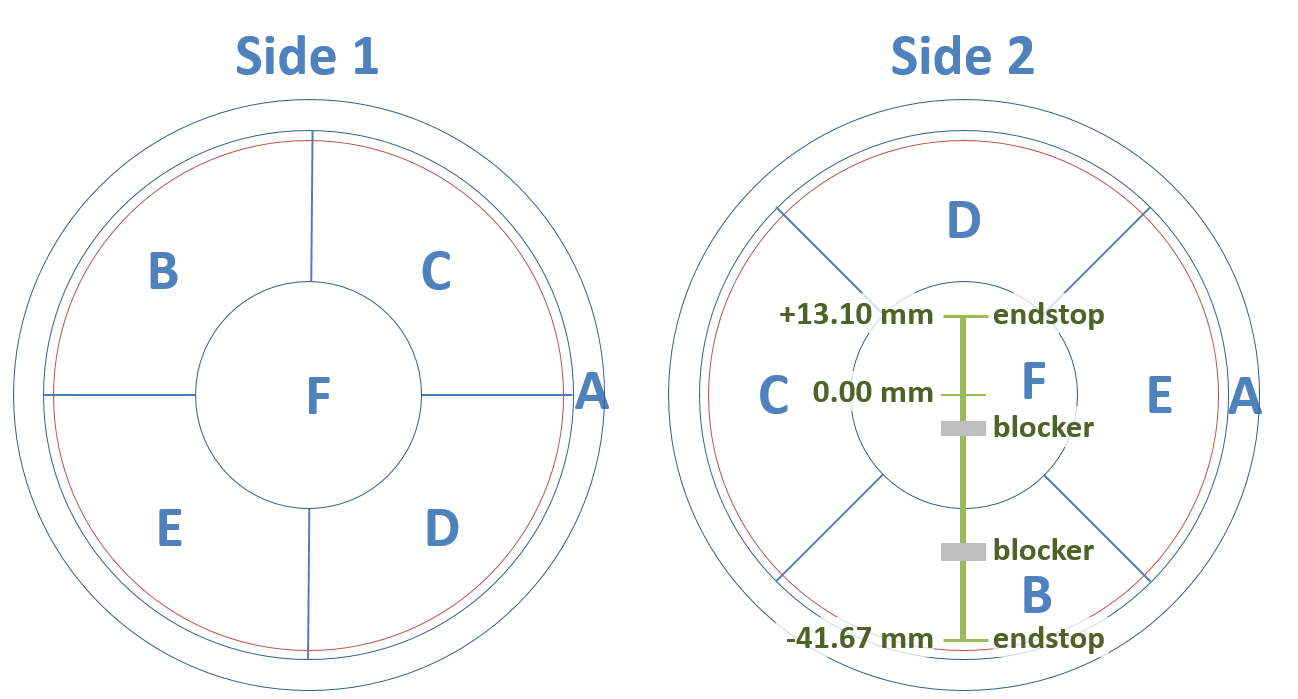}
\caption{\label{fig:izip}Detector layout and calibration source path. Divisions between phonon sensors (labeled A-F on each side) indicated in blue and between charge sensors (inner and outer on each side) in red. Source range of motion indicated in green.}
\end{figure}

The motor and source assembly was mounted on a copper plate with a slot through which the gammas could pass unimpeded. This plate was mounted in a copper housing \SI{1.6}{\centi\meter} above the particle detector, allowing the calibration source to illuminate one face of the detector. The source was moved along the detector diameter as shown in Fig.~\ref{fig:izip}.

Two endstops were used to validate the source position. At either end of the source holder's range of motion, we mounted insulated rigid copper strips connected to wires leading out of the fridge. These were arranged such that the source holder would come into contact with them at the extreme positions of its motion. Contact was identified by monitoring the resistance between these contacts and ground. 

In order to confirm the position within its range, two small Indium blockers (\SI{1}{\milli\meter} thick, \SI{\sim 2}{\milli\meter} wide) were inserted in the slot of the copper plate, thus attenuating the source gammas at those locations. Such blockers also provided a means to effectively toggle the source illumination off and on.

The source mover assembly was mounted above a prototype SuperCDMS iZIP detector \cite{Chagani14}. The iZIP is a cylindrical germanium crystal \SI{100}{\milli\meter} in diameter and \SI{33.3}{\milli\meter} thick. Each face of the device is patterned with 6 phonon sensors and 2 charge electrodes as shown in Fig.~\ref{fig:izip}. The phonon sensors are photolithographically patterned Quasi-particle trap assisted Electrothermal feedback Transition edge sensors (QETs) \cite{Irwin2005}. The phonon and charge sensors are interleaved in order to identify events near the crystal surfaces. 

For this study, the detector was operated with biases of \SI{\pm4}{\volt} on the two faces. The detector was read out using existing CDMS-II cold hardware \cite{Akerib08} with minor modifications, using prototype SuperCDMS detector control and readout cards \cite{Hansen10}.

\section{\label{sec:operation}Operation}
The detector and movable source were mounted inside an Oxford Instruments Kelvinox 100 dilution refrigerator \cite{Kelvinox} and cooled to \SI{\sim 35}{\milli\kelvin}. 
Excess heating of the base temperature stage was observed when operating the stepper motor.

The likely source was Joule heating in the connection between the Nb-Ti coil and wiring harness. textcolor{red}{Nb-Ti wire with no normally-conductive cladding (e.g. Cu, CuNi, PhBr) was used because it was available in the lab. However, bare Nb-Ti wire is notoriously difficult to solder and we fabricated the} joint by removing the Formvar insulation from the end of the Nb-Ti wire and tightly wrapping this within the strands of an aluminum pigtail wire. This was then covered with standard Pb-Sn solder and the joint was further secured with heat-shrink tubing. Due to the high normal resistance of the Nb-Ti wire, it is difficult to measure the resistance of these connections, but bench measurements indicate they may be $\lesssim$ \SI{3}{\ohm}. These joints were then wrapped into the coil assembly with Teflon tape and packed in the motor housing when the motor was assembled. This arrangement, while protecting the delicate Nb-Ti wires from damage, left the joints poorly heatsunk.

By sending low DC current (\SI{<5}{\milli\amp}) down a single coil, we observed fridge heating consistent with a resistance of $\mathcal{O}(\SI{1}{\ohm})$ at the location of the Nb-Ti joint and wiring harness. When driving \SI{\sim 35}{\milli\amp} through the coils to operate the motor, this $\mathcal{O}(\SI{1}{\milli\watt})$ of heating was strong enough to drive a portion of the adjoining Nb-Ti coil normal. Once this process began, the normal portion of the Nb-Ti would also heat, leading to a runaway increase in the coil resistance and heating. By monitoring the voltage and current required to drive the motor coils, we observed such increases in resistance when operating the motor for more than \SI{\sim 200}{\micro\second}. Operating the device for much longer periods of time would overwhelm the fridge with the substantial Joule heating. This made it difficult to estimate whether friction or eddy currents caused significant additional heating.

Even with these issues, we were able to accomplish reasonable movement of the source. Heating was minimized by operating with a low drive current of \SI{\sim 35}{\milli\amp} and only energizing the coils immediately before stepping the motor, then de-energizing them once the stepping was completed. Additionally the motor was only moved in bursts of \numrange{4}{6} steps at a time, since after this the coils heated and the drive current could not be maintained. 

Smoothing capacitors were added in parallel with each coil to reduce possible eddy heating due to transient spikes in the drive current signal output by the controller board. The time scale of this smoothing required that the dwell between step signals be \SI{40}{\milli\second}. While a shorter smoothing timescale would have allowed more steps before the coils heated, it may have also increased heating from transients, but this was not studied in detail.

When operated as described above, the base temperature would momentarily spike to \SI{\sim 500}{\milli\kelvin}, but return to \SI{< 50}{\milli\kelvin} within 5 minutes. By automating repeated bursts of movement and cooling, the motor could be operated at \SI{\sim 1}{step\per\minute} for arbitrary periods of time without overwhelming the refrigerator. While not ideal, this was fast enough to allow a cursory study of the operation of this system with the source at a variety of positions in a reasonable period of time. In this mode the source was translated the full distance between endstops. The number of step commands sent to the motor was consistent with the number required to move this distance, implying 100\% step efficiency. This allowed us to estimate source locations to within \SI{60}{\micro\meter}.
\section{\label{sec:pce}Phonon Collection Efficiency}
It was discovered that, after each operation of the motor, the phonon collection efficiency (PCE) of the QETs was diminished. PCE is the efficiency with which phonons from the detector crystal are coupled to the transition edge sensor of the QET. We believe this was a result of the superconducting Al fins trapping magnetic flux from the motor drive coils. This flux in turn trapped diffusing quasiparticles in the Al, reducing the fraction which heated the TES and thus lowering the observed signal \cite{Brink}.

\begin{figure}[h]
\includegraphics[width=\columnwidth]{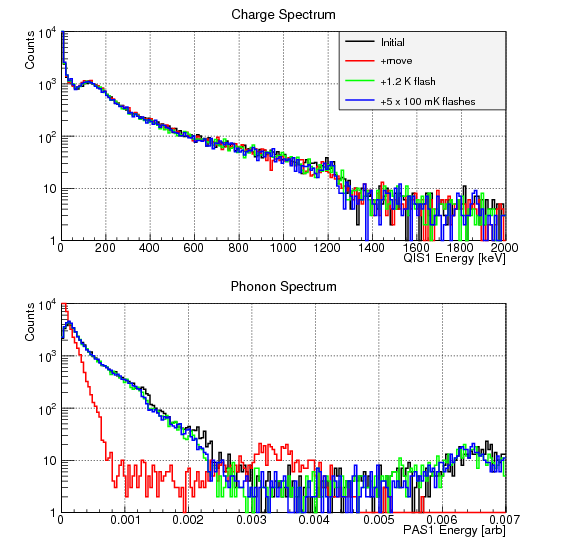}
\caption{\label{fig:pce} Charge and phonon spectra showing reduced PCE immediately after motor operation (+move) and restored PCE after \SI{1.2}{\kelvin} heating (+1.2 K flash). Subsequent LED heating and detector operation (+5 x 100 mK flashes) did not affect PCE.}
\end{figure}

After operating the motor, we observed reduced signal amplitudes on all phonon channels, but unchanged charge signals. This is demonstrated in Fig.~\ref{fig:pce} and indicates lower PCE at the phonon sensors as opposed to charge trapping or other crystal effects. Operating the motor heated everything at the base temperature stage (including the detector) and produced magnetic fields outside of the motor housing of $\mathcal{O}( \text{\SI{100}{\micro\tesla}})$. This heating was likely sufficient to drive the Al fins normal such that, when they cooled again, they trapped magnetic flux still present from the motor. The maximum temperature measured at the motor mounting plate was \SI{\sim 500}{\milli\kelvin}, but the maximum temperature of the aluminum fins could be higher; the fins are exposed to direct IR radiation from the motor, and there is a delay of some 10s of seconds until the detector temperature comes into equilibrium with the base temperature stage through the Cirlex clamps that hold the detector in its housing.

We were able to restore the PCE to its original value by heating the detector above the critical temperature of the Al fins (\SI{1.2}{\kelvin}) for a few seconds using LEDs installed inside the detector housing. The PCE was unchanged after subsequent cycles of LED heating and detector operation, indicating that \SI{1.2}{\kelvin} heating fully restored PCE. We adopted a procedure of moving the source then briefly heating the detector above \SI{1.2}{\kelvin}, restoring PCE. After this heating, the detector would cool back to \SI{\sim 35}{\milli\kelvin} within 2 hours at which time the detector could be operated.
\section{\label{sec:phonon}Example Application: Phonon Timing}
An in-situ movable calibration source would allow SuperCDMS to study the position dependence of phonon signal arrival times. The charge transport and phonon physics in such devices is non-trivial, and, in combination with relevant simulations, timing data such as this can be used to constrain model parameters. 

\begin{figure}[h]
\includegraphics[width=\columnwidth]{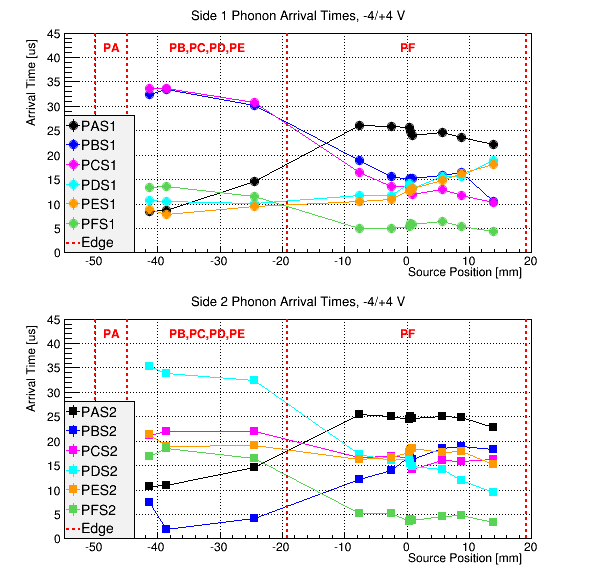}
\caption{\label{fig:pdelay} Phonon arrival times as a function of calibration source position. Vertical red lines and labels indicate the boundaries between channels along the detector diameter.}
\end{figure}

Fig.~\ref{fig:pdelay} shows the mean arrival times of phonon signals at the various sensors as a function of the calibration source position. Each phonon channel was labeled by channel identifier and side number. e.g. ``PAS1'' indicates phonon channel A on side 1 of the detector (see Fig.~\ref{fig:izip}). The arrival time was calculated as the delay between the beginning of the charge signal and the beginning of each phonon signal. In particular, the phonon signal start times were extrapolated from each channel's 10\% and 30\% phonon pulse rise times.

The detector employed for this study featured a design with interleaved charge and phonon sensors which allow the identification of events occurring within \SI{\sim 1}{\milli\meter} of the detector surface, where they suffer from reduced ionization yield \cite{Agnese13}. These surface events are identified in data as having very asymmetric charge signals between side 1 and side 2. The events selected to produce Fig.~\ref{fig:pdelay} were 60 keV events (as measured by both charge and phonon signal amplitudes) which did not occur near such a surface (as determined from the charge signal symmetry).

By comparing these arrival times to the channel locations it can be seen that phonon arrival times generally decrease as the source is moved closer to a particular channel, as expected. The effective phonon propagation speeds are also observed to be a few times slower than their sound speeds in Ge. This agrees with our understanding of phonon propagation in these devices as being a combination of diffusive and ballistic transport \cite{Hertel2011}. More detailed analysis of these trends requires modelling of the charge and phonon transport physics at each location in this detector's particular bias field arrangement and is beyond the scope of this paper. 

Data sets generated by using a movable source may be used to validate detector simulations and study models of charge and phonon transport. A set of detailed scans could also be used to fully map a detector's response throughout the detector volume. While the above example used 60 keV events occurring in the bulk of the detector, surface event behavior as a function of position can also be studied using lower energy ($\sim$10 keV) gammas or betas which do not penetrate as far into Ge.

\section{\label{sec:discussion}Discussion}
A simple movable source design such as this will be very useful for detailed detector characterization studies. It also provides a simple way to toggle an in-situ source off and on by placing it behind a blocker.

For the particular design tested in this study, solving the coil heating issues will enable faster source translation, though there will still be limits due to frictional heating. One can also trade longer pitch thread for increased speed at the loss of position accuracy. To address the Joule heating a different connection method using superconducting Al screw terminals is being explored, but this could be entirely avoided by using Nb-Ti clad with a normal conductor instead. The inclusion of a small mu-metal shield around the motor could fix the problems associated with stray magnetic fields. Future modifications are planned in order to add an angular degree of freedom with a secondary motor to enable a full two-dimensional scan of the detector face.

The authors would like to thank the SuperCDMS collaboration for providing the detector and electronics. We also thank Jake Nelson and Gabriel Spahn for their help with lab operations. This work was supported by DOE grant DE-SC0012294.









\bibliographystyle{bst/elsarticle-num}

\bibliography{bibliography}

\providecommand{\noopsort}[1]{}\providecommand{\singleletter}[1]{#1}%
\begin{thebibliography}{10}
\expandafter\ifx\csname url\endcsname\relax
  \def\url#1{\texttt{#1}}\fi
\expandafter\ifx\csname urlprefix\endcsname\relax\def\urlprefix{URL }\fi
\expandafter\ifx\csname href\endcsname\relax
  \def\href#1#2{#2} \def\path#1{#1}\fi

\bibitem{Agnese13}
R.~Agnese, et~al., {Demonstration of Surface Electron Rejection with
  Interleaved Germanium Detectors for Dark Matter Searches}, Applied Physics
  Letters 103 (2013) 164105.
\newblock \href {http://dx.doi.org/10.1063/1.4826093}
  {\path{doi:10.1063/1.4826093}}.

\bibitem{Broniatowski09}
A.~Broniatowski, et~al., A new high-background-rejection dark matter ge
  cryogenic detector, Physics Letters B 681~(4) (2009) 305 -- 309.
\newblock \href {http://dx.doi.org/10.1016/j.physletb.2009.10.036}
  {\path{doi:10.1016/j.physletb.2009.10.036}}.

\bibitem{Porter94}
F.~Porter, et~al., {A stepper motor for use at temperatures down to 20 mK},
  Physica B: Condensed Matter 194-196 (1994) 151 -- 152.
\newblock \href {http://dx.doi.org/10.1016/0921-4526(94)90405-7}
  {\path{doi:10.1016/0921-4526(94)90405-7}}.

\bibitem{EasyDriver}
B.~Schmalz, \href{http://www.schmalzhaus.com/EasyDriver/}{{EasyDriver Stepper
  Motor Driver, v4.5}}.
\newline\urlprefix\url{http://www.schmalzhaus.com/EasyDriver/}

\bibitem{Arduino}
\href{https://www.arduino.cc/}{{Arduino AG}}.
\newline\urlprefix\url{https://www.arduino.cc/}

\bibitem{Chagani14}
H.~Chagani, et~al., {First Measurements of SuperCDMS SNOLAB 100 mm Diameter
  Germanium Dark Matter Detectors with Interleaved Charge and Phonon Channels},
  Proceedings of Science 161.
\newblock \href {http://dx.doi.org/10.22323/1.213.0161}
  {\path{doi:10.22323/1.213.0161}}.

\bibitem{Irwin2005}
K.~Irwin, G.~Hilton, {Transition-Edge Sensors}, in: Cryogenic Particle
  Detection, Springer, 2005, pp. 63--150.
\newblock \href {http://dx.doi.org/10.1007/10933596_3}
  {\path{doi:10.1007/10933596_3}}.

\bibitem{Akerib08}
D.~Akerib, et~al., {Design and performance of a modular low-radioactivity
  readout system for cryogenic detectors in the CDMS experiment}, Nuclear
  Instruments and Methods A 591 (2008) 476 -- 489.
\newblock \href {http://dx.doi.org/10.1016/j.nima.2008.03.103}
  {\path{doi:10.1016/j.nima.2008.03.103}}.

\bibitem{Hansen10}
S.~{Hansen}, et~al., The cryogenic dark matter search test stand warm
  electronics card, in: IEEE Nuclear Science Symposuim Medical Imaging
  Conference, 2010, pp. 1392--1395.
\newblock \href {http://dx.doi.org/10.1109/NSSMIC.2010.5874000}
  {\path{doi:10.1109/NSSMIC.2010.5874000}}.

\bibitem{Kelvinox}
{Oxford Instruments}, {Tubney Woods, Abingdon, Oxfordshire OX13 5QX, United
  Kingdom}.

\bibitem{Brink}
P.~Brink, {Non-Equilibrium Superconductivity induced by X-ray Photons}, Ph.D.
  thesis, Magdalen College, Oxford (1995).

\bibitem{Hertel2011}
S.~A. Hertel, M.~Pyle, {Phonon Pulse Shape Discrimination in SuperCDMS Soudan},
  J. Low. Temp. Phys. 167 (2012) 1173--1178.
\newblock \href {http://dx.doi.org/10.1007/s10909-012-0498-6}
  {\path{doi:10.1007/s10909-012-0498-6}}.

\end{thebibliography}

\end{document}